%% file: main.tex
\documentclass[twocolumn,9pt]{article}
\usepackage[square,sort,comma,numbers]{natbib}
\usepackage[utf8]{inputenc}
\usepackage[margin=0.9in,paper=letterpaper]{geometry}

\usepackage{paralist} 
\usepackage{outlines}
\usepackage{multicol}
\usepackage{booktabs} 
\usepackage{amssymb}  
\usepackage{graphicx} 
\usepackage{makecell}
\usepackage{wrapfig}
\usepackage{lscape}
\usepackage{appendix}
\usepackage[most]{tcolorbox} 
\usepackage{subfigure} 
\usepackage{tikz,pgfplots,pgfplotstable} 
\usepackage{mdframed} 
\pgfplotsset{compat=1.7} 
\usepackage{pifont} 
\usepackage{amsmath} 
\usepackage{bm} 
\usepackage{multirow} 
\usepackage{enumitem} 

\usepackage{url}

\usepackage{libertine}
\usepackage{libertinust1math}
\usepackage[T1]{fontenc}

\begin{document}

\title{\large \bf Widely Reused and Shared, Infrequently Updated, and Sometimes Inherited: A Holistic View of PIN Authentication in Digital Lives and Beyond\thanks{A version of this paper is appears at the 2020 Annual Computer Security Applications Conference (ACSAC'20).}}

\author{Hassan Khan\\
{\small University of Guelph}\\
{\small \texttt{hassan.khan@uoguelph.ca}}\\
\and
Jason Ceci\\
{\small University of Guelph}\\
{\small \texttt{jceci@uoguelph.ca}}\\
\and
Jonah Stegman\\
{\small University of Guelph}\\
{\small \texttt{jstegman@uoguelph.ca}}\\
\and
Adam J. Aviv\\
{\small The George Washington University}\\
{\small \texttt{aaviv@gwu.edu}}\\
\and
Rozita Dara\\
{\small University of Guelph}\\
{\small \texttt{drozita@uoguelph.ca}}\\
\and
Ravi Kuber\\
{\small University of Maryland, Baltimore County}\\
{\small \texttt{rkuber@umbc.edu}}}

\date{}


\newcommand{\ravi}[1]{\textcolor{magenta}{[\textit{Ravi: #1}]}}
\newcommand{\adam}[1]{\textcolor{orange}{[\textit{Adam: #1}]}}
\newcommand{\hassan}[1]{\textcolor{blue}{[\textbf{Hassan: #1]}}}
\newcommand{\harsh}[1]{\textcolor{green}{[\textbf{Harsh: #1]}}}

\newcommand{\new}[1]{\textcolor{black}{#1}}

\def\changemargin#1#2{\vspace{-2pt}\list{}{\rightmargin#2\leftmargin#1}\item[]}
\def\endchangemargin {\endlist \vspace{-2pt}}

\maketitle

\input{00-abstract}
\input{01-introduction}
\input{02-related_work}
\input{03-study}

\input{04-results}
\input{05-discussion}

\input{06-future_work}

\input{07-limitations}

\input{08-conclusion}

\section*{Acknowledgments}
We thank Flynn Wolf, Harshvardhan Verma, and Kassidy Marsh for their feedback on the survey and assistance. This material is based upon work supported by NSERC under Grant No. RGPIN-2019-05120 and the National Science Foundation under Grants No. 1845300. Any opinions, findings, and conclusions or recommendations expressed in this material are those of the authors and do not necessarily reflect the views of the funding agencies.

\bibliographystyle{plain}
\bibliography{references}


\appendix
\input{99-appendix}

\end{document}

%% file: 00-abstract.tex
\begin{abstract}
    Personal Identification Numbers (PINs) are widely used as an access control mechanism for digital assets (e.g., smartphones), financial assets (e.g., ATM cards), and physical assets (e.g., locks for garage doors or homes).
    Using semi-structured interviews (n=35), participants reported on PIN usage for different types of assets, including how users choose, share, inherit, and reuse PINs, as well as behaviour following the compromise of a PIN. We find that memorability is the most important criterion when choosing a PIN, more so than security or concerns of reuse. Updating or changing a PIN is very uncommon, even when a PIN is compromised. 
  Participants reported sharing PINs for one type of asset with acquaintances but inadvertently reused them for other assets, thereby subjecting themselves to potential risks.
 Participants also reported using PINs originally set by previous homeowners for physical devices (e.g., alarm or keypad door entry systems).  While aware of the risks of not updating PINs, this did not always deter participants from using inherited PINs, as they were often missing instructions on how to update them. 
 Given the expected increase in PIN-protected assets (e.g., loyalty cards, smart locks, and web apps), we provide suggestions and future research directions to better support users with multiple digital and non-digital assets and more secure human-device interaction when utilizing PINs.
  
\end{abstract}


%% file: 01-introduction.tex

\section{Introduction}
\label{sec:introduction}

Knowledge-based authentication (e.g., passwords or PINs) is widely used as it is a well-tested technology and simple to administer~\cite{renaud2004my}. 
However, research suggests that there are persistent challenges with password usability~\cite{bonneau2012quest} and memorability~\cite{dhamija2000deja, garfinkel2014usable}. 
Additionally, passwords are too cumbersome to use for protecting certain classes of assets, such as a car or garage door, which limits their utility.

With the proliferation of technology, it is somewhat ironic that passwords ``stubbornly survive and reproduce with every new website''~\cite{bonneau2012quest}.  Challenges with passwords can lead to frustration among users. 
To address these lingering concerns, several mobile and web apps now provide PIN-based authentication as the default option~\cite{paypal_pin}. 
Microsoft is also planning to remove the password option from the Windows 10 login screen while keeping PIN as one of the login options~\cite{win10pin}. 
Loyalty cards also require PINs to redeem points, and a survey indicates that an average Canadian participates in twelve loyalty programs, which is a 25\% increase over four years~\cite{macorr_loyalty}. 
Keyless home locks require PINs to authenticate, and their market is forecasted to reach 35 million units by 2027~\cite{grand_view_smartlock}.
As technologies requiring security in the form of PINs become more prevalent, it is critical to understand how people choose and manage PINs, not just for digital and financial assets, but for the wide array of physical assets for which PINs are used. 

In studying PIN management, we broadly categorize PINs into three categories of protected assets: digital (e.g., to unlock digital devices or authenticate to mobile and web apps), financial (e.g., ATM cards or banking apps), and physical (e.g., digital keypad based entry systems for garages or homes).
Researchers have explored PIN-based authentication for financial assets, notably Bonneau et al. studied chip-and-PIN systems~\cite{bonneau2012birthday}, as well as Wang et al. studied the guessability of PINs as derived from leaked password  datasets~\cite{wang2017understanding}. 

We argue that a broader analysis of PIN usage needs consideration for several reasons. First, different types of assets may be subject to different types of attacks (e.g., smartphone PINs might be more susceptible to shoulder surfing~\cite{deluca2010atm-sec} than PINs used to protect physical assets), and prevalent reuse across these categories may result in undesirable consequences and increased risks.
Second, PINs for certain types of assets may be more likely to be shared (e.g., financial vs. digital asset PINs with a family member), and their careless reuse may result in unauthorized usage.
Third, physical PINs are more likely to be shared among family members or by trusted individuals within their network, which leads to interesting issues surrounding selection of PINs. 

Prior work has yet to focus deeply on ways to address the broader considerations associated with PIN usage from a user perspective.  In this study, we aim to investigate how users create, manage, and share PINs across different types of assets. To this end, we conducted hour-long, semi-structured interviews with 35 participants.
We chose a semi-structured methodology to unpack and better understand the themes relating to use of PINs. 
Our findings include:

\begin{itemize} \itemsep-1pt
    \item When selecting a PIN, participants were more likely to prioritize memorability of the PIN over security.
    While participants reported that reusing a PIN was a low factor in selecting a PIN for a given asset, the majority of participants (28/35 or 80\%) reported reusing PINs.  
    This reuse was across different asset types and often resulted in PINs for physical devices (e.g., bike lock) moving into the digital world and vice versa.
    \item Despite more than two-thirds (71\%) of our participants describing situations where their PINs were compromised, less than half of those (45\%) reported updating their PINs. This can be attributed in part to concerns relating to memorability and usability of PINs.
    
    \item PIN update is very uncommon, overall, and when it does occur, it is often due to reasons of security or memorability (26/49 of reported PIN updates). However, for physical assets such as garage doors, a lack of update may be due to the nature of these devices. Six out of nine owners of PIN-protected garage doors reported that they were unable to update their PIN as they did not know how to perform this action, despite desiring to do so. 
    
    \item Differences between asset types influence the security measures adopted by users. Participants were less worried about compromising their physical PINs compared to digital PINs, as potential attackers breaking into an entity protected by a physical PIN may face criminal prosecution, e.g., breaking and entering, despite the fact that digital or financial PINs can also lead to personal, financial, or criminal harm.  
\end{itemize}

Based on our findings, we propose three areas for further exploration. First, new intervention and strategies for assisting users in selecting and recalling PINs would address many of the observed shortcomings. While password managers are an obvious solution, their usage is mostly focused on different types of accounts.  However, current password managers could be augmented to assist these tasks. Second, as PINs become more pervasive, 
users may become more concerned with the threat of shoulder surfing attacks. To counteract, 
the research community should focus on developing new tools to assist users in identifying instances of shoulder surfing, and provide guidance on mitigation practices. Finally, given the plethora of PIN usage scenarios, unifying methods for updating PINs, similar to how password changing has mostly stabilized around standard practice, would make a difference in encouraging PIN updates after compromise. Of course, for physical assets, this is not a simple task. Perhaps augmented reality tools could be used to address this gap in the future, to link these physical assists to known documentation. 



%% file: 02-related_work.tex
\section{Related Work}
\label{sec:related_work}

In this section, we explore related work in areas including: PIN choices for human-chosen PINs, attacks on PINs, memorability and reusability of PINs, and lifecycle and management of authentication credentials in general. We also compare and contrast our findings for specific topics related to PIN usage with findings for other authentication methods in Section~\ref{sec:discussion}.

\subsection{Human-Chosen PINs}
Users face several choices when choosing their authentication secrets. Selection is often influenced by factors such as memorability of the chosen secret, reuse of an existing secret, usability (including time to authenticate and error rates), and security~\cite{bonneau2012quest, chiasson2006usability, schechter2015learning}. 
Von Zezschwitz et al.~\cite{von2013survival} have explored users' choices for text-based password composition, while Biddle et al.~\cite{biddle2012graphical} have summarized research that explores users' choices of graphical passwords.
PINs are less complex than text-based passwords~\cite{jakobsson2011bootstrapping} and different from graphical passwords since PINs require memorizing digits. 

Amitay collected PINs surreptitiously from an iPhone app in the App Store. 
Their data showed that ten of the most commonly used 4-digit PINs represented 15\% of all PINs in use~\cite{amitay2011common}. Furthermore, most of these PINs followed simple patterns of repeating or consecutive digits.
In a seminal work, Bonneau et al.~\cite{bonneau2012birthday} explored the user selection preferences for bank card PINs (e.g., chip-and-PIN systems) using survey data and approximated PINs from leaked password data and Amitay's dataset. They found that an attacker who comes into the possession of a lost wallet with a bank card and owner's ID in it has about an 8\% chance of guessing the correct PIN due to the widespread use of birthdays for PINs. Wang et al.~\cite{wang2017understanding} compared characteristics (guessability, entropy, and distribution) of chosen 4-/6-digit PINs between English and Chinese users. Among other findings, they showed that the top 5-8\% most popular PINs account for over 50\% of PIN datasets.
Markert et al.~\cite{markert-20-pin-blacklist} collected data on 4-/6-digit PINs, also finding high prevalence of popular PINs, and that the benefit of using 6-digit PINs is minimal (or worse) than a 4-digit PIN.
\new{Concurrent to this research, Casimiro et al.~\cite{casimiro2020quest} conducted an MTurk survey to study PIN choices and reuse and confirm our findings.}
 While these studies offer an insight into the prevalent reuse and not-so-secret nature of human PIN choice, our research extends prior work by examining users' motivations behind their choices. \looseness=-1

\subsection{Attacks on PINs and Defences}
A range of studies have focused on the development and evaluation of novel interaction techniques to defend against shoulder surfing attacks~\cite{de2013back,de2014now,leiva2014bod, de2010colorpin, von2015swipin}.
Researchers have also explored novel side channel-based attacks on PIN authentication, but these attacks require special equipment or skillful attackers~\cite{abdelrahman2017stay, foo2010timing, xu2012taplogger}.
Since these efforts are only tangentially related to our work,  we discuss more related works that study attacks and the recourse of victims. 

Aviv et al.~\cite{aviv2017towards} and Khan et al.~\cite{khan2018evaluating}, empirically evaluated the success of shoulder surfing attacks on PINs under various conditions. 
De Luca et al.~\cite{deluca2010atm-sec} found that German ATM users reported a low incidence of PIN shielding during ATM use.
They also reported a significant influence
of factors such as distractions, physical hindrance, trust relationships, and memorability on security in PIN-based ATM use. 
Harbach et al.~\cite{harbach2014sa} conducted an online survey and field study to understand users' smartphone unlocking behaviour. Of users that use a lock code (including PIN and graphical pattern users) for their smartphones, 65\% were not or mostly not concerned about a shoulder surfing attack on their code. 
Other related work includes the study by Eiband et al.~\cite{eiband2017understanding}, who explored shoulder surfing attacks and defences during normal smartphone usage, without focusing on authentication.  

Our work expands the existing body of knowledge by exploring attacks on PINs, the defences that are employed, and the recourse of users when they suspect that the attacks are successful for various digital and non-digital assets.


\subsection{Security and Memorability of PINs} 
In an attempt to encourage users to be more secure in their authentication behaviour, researchers have explored methods to generate and help users memorize secure PINs.
Kim and Huh~\cite{kim2012pin} found that using a blacklist policy of restricting around 200 commonly used PINs significantly increases the randomness (as measured using Shannon entropy, not guessability~\cite{bonneau20120science}) of PINs without significantly increasing the memorability overhead. Findings from a study by Markert et al.~\cite{markert-20-pin-blacklist} indicate that even small blacklists of disallowed PINs can substantially improve the security (as measured using guessability) of user-chosen PINs against throttled attackers.
Schechter and Bonneau~\cite{schechter2015learning} proposed two techniques to memorize secure PINs and conducted a study to show that the proposed memorization techniques were effective, thereby reducing the likelihood of writing down the new PIN. 
Stanekova and Stanek~\cite{stanekova2013analysis} and Huh et al.~\cite{huh2015memorability} also explored effective methods to generate and memorize PINs.
Our work explores memorability and usage issues surrounding PINs without exploring users' memorization strategies, and our findings provide further motivation for the development of effective PIN memorization techniques.

Renaud and Volkamer~\cite{renaud2015management} conducted an online study to evaluate two PIN memorization assistance techniques. While they reported no improvements in PIN memorization 
due to the users not using the memorization aids, they reported on the strategies people adopted for PIN memorization and whether participants wrote down their PINs. They also identified reasons why participants updated their PINs. However, they did not specify the rate at which different PIN changes occurred and for what reason. We conduct a more holistic and broader investigation of these phenomena.
We categorize and quantify the reasons why participants change PINs and report on instances when participants chose not to change their PINs after PIN compromise for different asset categories.

\subsection{Lifecycle of Authentication Credentials}

Although the lifecycle and management of PINs have not been subjects of much research (either in digital or non-digital contexts), researchers have explored these topics for passwords. 
Stobert and Biddle~\cite{stobert2014password} investigated how users managed their passwords through a series of interviews. 
They reported that users ration their efforts to protect their accounts best, and many users reuse passwords as well as adjust them for different accounts. They also found that people were willing to put more effort into the management of accounts with higher perceived importance (i.e., bank account passwords). 
Hayashi and Hong~\cite{hayashi2011diary} conducted a two-week diary study to examine password usage of 20 users.  
They collected data on the frequency and location of password use, and the use of password aids. Based on their findings, they provide suggestions to improve the password authentication experiences of users.

As PIN-based authentication increasingly becomes one of the default authentication options for digital, physical, and financial assets, it is important to understand PIN lifecycle and management across different assets.
Our study is the first of its kind to report a holistic view of the lifecycle and management of PINs, thereby highlighting interrelationships across PINs for different types of assets. \looseness=-1

%% file: 03-study.tex
\section{Study Design and Methodology}
\label{sec:study}

\paragraph{Design}
The aim of our study is to better understand how individuals use PINs across a variety of assets.
However, there are several challenges to such holistic explorations.
First, users may not be attentive to how their PIN management behaviour varies across different assets.
Therefore, we chose to conduct semi-structured interviews, which allowed participants to speak openly about their PIN management experiences.
This format also  provides us with quantitative data, as well as enabled us to ask clarifying and follow-up questions in cases where more detail is needed for qualitative analysis. \looseness=-1

Second, collecting user-selected PINs, as used to access a wide range of assets, can quickly become impractical due to the many-to-many mapping between PINs that users employ and the different asset categories. Such a study, while valuable, would be incredibly time consuming and perhaps an error-prone task. Using semi-structured interviews allows us to perform exploratory analysis on the topic with respect to the types of assets protected by PINs and to investigate usage strategies.  The findings would provide a greater awareness of exact PINs used in each asset class. 

In developing our survey instrument, we initially conducted a pilot study (n = 4) with participants from the first author's department. These participants were invited to a lab where they undertook a structured survey containing questions related to their demographics, their self-reported proficiency with technology and computer security, and whom they lived with (see Appendix~\ref{app:demo} for details).
We then asked participants to enumerate all the PINs that they use, and then for each PIN, we inquired about selection (when and how they went about choosing it), resources it protects and the perceived sensitivity of each resource.  We then asked about the frequency of PIN entry, others whom they shared that PIN with, and the perceived  trust  in those individuals. We also asked about any attacks that had been encountered and their recourse.
Finally, we asked participants questions applicable to all categories, including sharing across categories, and PIN management after they moved on from a relationship where they had shared a PIN with another individual. \looseness=-1

During the pilot, participants had to respond to the same set of questions for up to seven PINs. As a result, they found the survey instrument to be cumbersome, as some of the questions felt unnecessarily repetitive. 
We addressed this by redesigning our survey across four sections: a section that contained questions that were independent of any asset category or were pertaining to all asset categories; and three sections that contained the same set of questions for each of the three asset categories.
This enabled us to collect qualitative data effectively for each PIN category without fatiguing the participants. 

From the pilot study, we also noted that users were using multiple PINs in each category (e.g., multiple PINs for multiple banking cards).
In the updated survey, while we collected information on how many PINs participants used for each category and across how many assets, we asked participants to respond to our category-specific questions (i.e., PIN choice sharing, reuse, and security-related aspects) for the most used PIN in each category. 
While this design choice may have resulted in losing some valuable information, it also supported our objective of collecting high-quality data without losing participants' interest due to unnecessary repetition. \looseness=-1

The redesigned survey was conducted with a new group of pilot participants (n = 4). 
All participants completed the surveys within an hour, a more acceptable time frame. 
The researchers examined the data and found that the categorization of questions across different categories provided more meaningful insights into participants' behaviour regarding the PIN lifecycle. 
Therefore, this improved survey was employed for our main study (also provided in Appendix~\ref{app:semi}).


  %

\paragraph{Methodology}
We received approval from our ethics board for this study.
We recruited participants from a local classified ad portal, flyers posted around the local area, and word-of-mouth advertising. 
Participants were offered \$25 for their participation in an hour-long study conducted on campus at the University of Guelph. 
They were informed prior to participating in the interview, that they must not reveal their actual PINs to the researchers. 

\begin{table}[t]
\caption{Participants' demographics (\textsuperscript{*}UD~=~Undisclosed)}
\label{tab:demographics}
\centering
\small\addtolength{\tabcolsep}{-3.75pt} \renewcommand{\arraystretch}{1.05}
\begin{tabular}{cccccccc}
\multicolumn{8}{c}{{\bf n = 35}} \\ \hline \hline
\multicolumn{8}{c}{{\bf Gender} } \\
\multicolumn{4}{c}{\emph{Female}}  & \multicolumn{4}{c}{\emph{Male}} \\
\multicolumn{4}{c}{17}    & \multicolumn{4}{c}{18} \\ \hline
\multicolumn{8}{c}{{\bf Age (in years)}} \\
\emph{18--25}               & \emph{26--30}               & \emph{31--35}               & \emph{36--40}      & \emph{41--45}               & \emph{46--50}               & \multicolumn{2}{c}{\emph{50+}}                               \\
8                 & 4                 & 5                 & 6                 & 6                 & 1                  & \multicolumn{2}{c}{5}                                \\ \hline 
\multicolumn{8}{c}{{\bf Annual Household Income ($\times$ \$1000)} }  \\
\emph{\textgreater{}\$15}   & \emph{\$15--29}             & \emph{\$30--49}             & \emph{\$50--74}             & \emph{\$75-99}              & \emph{\$100--150}           & \emph{\textgreater{}\$150}  & \emph{UD\textsuperscript{*}} \\
 2 & 2  & 3  & 4 & 5 & 10 & 2 & 7                                 \\ \hline
\multicolumn{8}{c}{{\bf Highest Education Level}} \\
\multicolumn{3}{c}{\emph{High School}}                                    & \multicolumn{3}{c}{\emph{Undergraduate}}                                  & \multicolumn{2}{c}{\emph{Graduate}}                           \\
\multicolumn{3}{c}{17}                                           & \multicolumn{3}{c}{6}                                           & \multicolumn{2}{c}{12}                               \\ \hline
\multicolumn{8}{c}{{\bf Self Reported Proficiency in Technology}} \\
\multicolumn{3}{c}{\emph{Basic}}                                         & \multicolumn{3}{c}{\emph{Intermediate}}                                   & \multicolumn{2}{c}{\emph{Advanced}}                            \\
\multicolumn{3}{c}{6}                                           & \multicolumn{3}{c}{18}                                           & \multicolumn{2}{c}{11}                               \\ \hline
\multicolumn{8}{c}{{\bf Self Reported Proficiency in Security}} \\
\multicolumn{3}{c}{\emph{Basic}}                                          & \multicolumn{3}{c}{\emph{Intermediate}}                                   & \multicolumn{2}{c}{\emph{Advanced}}                            \\
\multicolumn{3}{c}{19}                                           & \multicolumn{3}{c}{9}                                           & \multicolumn{2}{c}{7}                            \\  \hline \hline
\end{tabular}

\end{table}

Before the interview, we described digital asset PINs as the PINs that are used to unlock digital devices or authenticate to mobile and web apps. 
Digital assets enumerated to participants included smartphones, laptops, personal computers, online accounts, voicemail, gaming consoles, apps, smart watches, thermostats, and other smart home devices.
While PINs to digital home locks or banking web or mobile apps could be classified as digital PINs, we asked participants to categorize those as physical or financial PINs, respectively.
Financial asset PINs were described as the PINs that controlled access to financial assets, including ATM cards, loyalty cards, and banking websites or apps.
Physical asset PINs were described as the PINs that controlled access to physical assets, including electronic home locks, home security systems, garage door openers, cars, and bike or gym locks.

During the semi-structured interview, the researcher first asked about the number of PINs participants used and the assets protected by these PINs.
The researcher also reminded participants about several assets that could be PIN protected to ensure that participants did not forget any PINs.
The researcher then explained each of the three categories of PINs, and provided examples of assets for each category.
The researcher then asked category-specific and category-independent semi-structured interview questions.

Table~\ref{tab:demographics} provides the demographic information of 35 participants and shows their diversity in terms of age, socio-economic group, education, and level of technology awareness.


%% file: 04-results.tex
\section{Results}
\label{sec:results}



    
    

\begin{figure}[t]
    \centering
    \includegraphics[width=64mm, trim={0 4mm 0 4mm},clip]{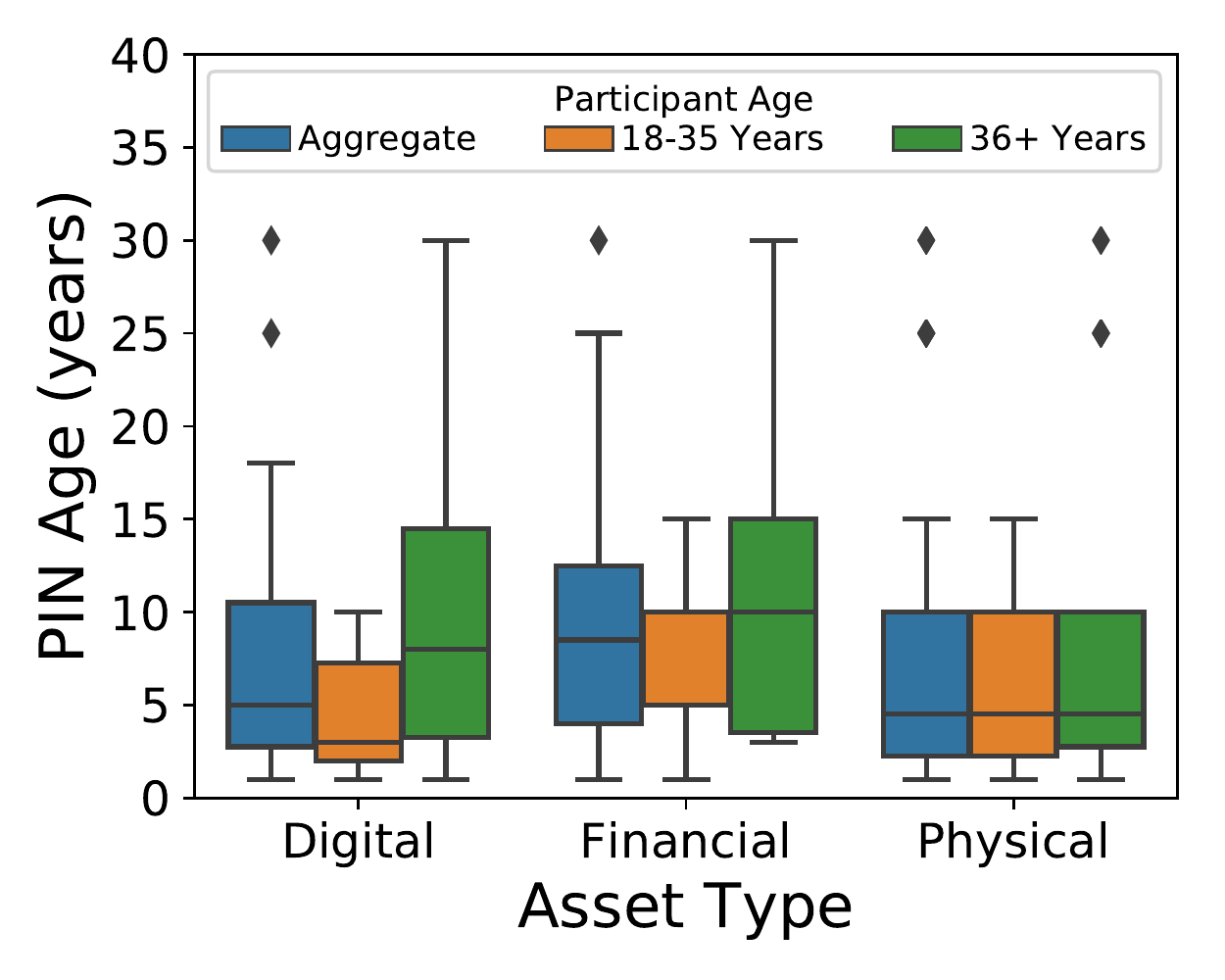}
    \caption{Boxplot of age of oldest PIN currently in use.}~\label{fig:pin_age}
  \end{figure}
  

\begin{table}[t]
  \caption{Statistics of 231 assets that were PIN-protected}~\label{tab:pin_stats}
  \centering
  \small 
  \begin{tabular}{l | c c c c c}
    {\small\textit{Asset}}
    & {\small \textit{Total}}
    & {\small \textit{Mean}}
      & {\small \textit{Median}}
    & {\small \textit{Min}}
    & {\small \textit{Max}} \\
    \hline
    Digital & 94 & 2.7 & 2 & 0 & 7\\
    Financial & 84 & 2.4 & 2 & 1 & 7\\
    Physical & 53 & 1.5 & 2 & 0 & 4\\
  \end{tabular}

\end{table}

We now present our findings.
For test statistics, a Pearson’s Chi-Squared test was used to compare categorical data, and a Kruskal-Wallis one-way analysis of variance was used to compare Likert scale responses between asset groups~\cite{kruskal1952use}. For all tests, a $p$ < $0.05$ critical value was used for statistical significance. For multiple comparisons of the same data category, we applied Bonferroni correction to $p$-values and set the significance cut-off at $\alpha/n$, where $n$ is the number of multiple comparisons~\cite{holm1979simple}. 
When reporting quotes from participants to represent a theme, we identify the number of participants who expressed that code and provide a representative quote.


\subsection{PIN Usage}
In total, 140 PINs were reportedly being used by our participants, and per participant, the average number of PINs in use is 4, with a minimum of 1 and a maximum of 15.
These PINs are used to control access to 231 assets.
As presented in Table~\ref{tab:pin_stats}, 94 (41\%) assets are digital, 84 (36\%) are financial, and 53 (23\%) are physical. \looseness=-1 


\begin{table}[t]
\caption{Reported PIN entry frequency across asset classes.}
\label{tab:pin_daily_use}
\centering
\small\addtolength{\tabcolsep}{-3.75pt} \renewcommand{\arraystretch}{1.05}
\begin{tabular}{l|ccc}
\textit{Frequency}     & \textit{Digital} & \textit{Financial} & \textit{Physical} \\ \hline
Multiple times/day   & 21/32 (66\%)     & 3/34 (9\%)         & 10/25 (40\%)      \\
Daily                  & 9/32 (28\%)      & 11/34 (32\%)       & 2/25 (8\%)        \\
Multiple times/week  & ---              & 10/34 (29\%)       & 9/25 (36\%)       \\
Weekly                 & 2/32 (6\%)       & 5/34 (15\%)        & 2/25 (8\%)        \\
Multiple times/month &  ---             & 1/34 (3\%)         & 1/25 (4\%)        \\
Monthly                &    ---           & 4/34 (12\%)        & 1/25 (4\%)       
\end{tabular}

\end{table}







Among the digital assets, participants primarily reported PINs for securing their smartphones, voicemail accounts, and laptops/PCs (32, 22, and 17 digital assets, respectively). 
For financial assets, participants reported using PINs for banking (debit or credit cards) and other loyalty cards (66 and 16 financial assets, respectively). 
Among physical assets, participants reported using PINs for keypad entry systems for  home (or security systems), garage doors, and dial locks for bikes/gym lockers (17, 19, and 11, respectively). 

Participants were asked to rate how important the \emph{security of their assets} is to them on a scale of 1--5 (5 being the most important) for each of the asset types.
The median response was 5 all asset types. The mean responses were 4.31, 4.71 and 4.23 for digital, financial and physical assets, respectively. A Kruskal-Wallis test 
indicated no significant differences between asset groups for the security rating ($H(2) = 4.98, p~=~0.08$). 

The self-reported daily usage of PINs across each category is provided in Table~\ref{tab:pin_daily_use}. 
Participants authenticated to their digital assets more frequently than financial or physical assets, 30/32 (94\%) ``Daily'' or ``Multiple times a day'' vs. 14/34 (41\%) and 12/25 (48\%, respectively.
While more participants were using their PIN-protected financial assets (e.g., bank cards) daily, they reported using more usable methods of payment, such as NFC-based tap-to-pay.

We asked participants to report the current PIN that they have been using for the longest period of time within each category.
Figure~\ref{fig:pin_age} shows the responses from all participants as well as responses grouped into two age groups---18--35 years (n = 16) and 36+ years (n = 19). For all participants, the median age of PINs for digital, financial, and physical assets was 5, 8.5, and 4.5 years, respectively. 
Six participants reported never changing a PIN across any category since configuring those.
As the sampled PIN ages were not distributed normally, a non-parametric Kruskal-Wallis one-way analysis of variance test was used to compare PIN ages between groups. 
However, this test provided no evidence to suggest that PIN age varied significantly between asset types ($H(2) = 2.94, p~=~0.23$). 

\begin{figure*}[t]
  \centering
  \includegraphics[width=1.8\columnwidth , trim={0 3mm 0 2mm},clip]{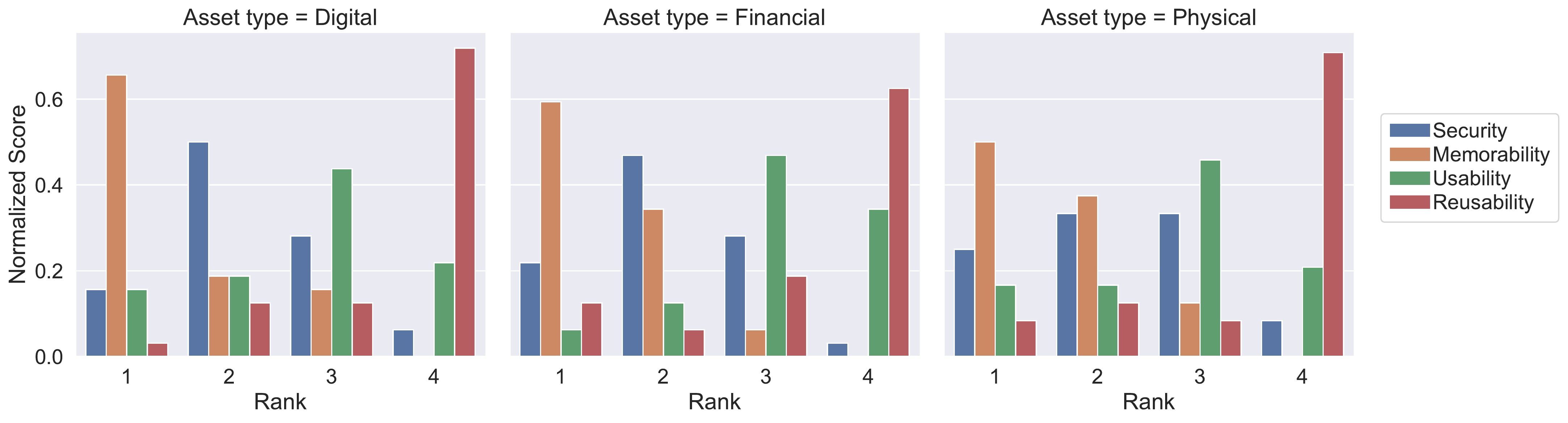}
  \caption{Participants' ranking of security, memorability, usability, and reusability criterion for PINs choices in different categories (lower rank indicates more important choice factor).}~\label{fig:pin_selection_criteria}
\end{figure*}

\subsection{PIN Choices}
\label{subsec:pin_choices}
We investigate the factors that motivate PIN choices by asking participants to rank the importance of four criteria when they are choosing PINs: security, memorability, usability, and reusability. 
The normalized score (rescaled to have values between 0 and 1) from the participants is plotted in Figure~\ref{fig:pin_selection_criteria}.
The ranking was normalized for better comparisons between PIN choices of different asset types.

 Figure~\ref{fig:pin_selection_criteria} shows that memorability is the most important factor for participants when they are choosing PINs across different asset types.
 Security and usability (defined as "ease to enter the PIN" for our participants to differentiate from memorability) were the next most important factors for the participants. While participants reported reusing PINs (see Section~\ref{subsec:pin_reuse}), they ranked reusability as the least important factor for different asset types. 
 The average ranks (1--4, 1 being most important) for memorability, security, usability, and reusability across assets were 1.52, 2.25, 2.83, and 3.40, respectively. A Kruskal-Wallis test shows a statistically significant difference between the ranks chosen for the four criteria ($H(3) = 130.93, p~<~0.01$). Post hoc pair-wise comparisons using Mann-Whitney U tests (Bonferroni corrected) between ranks given by participants for each criterion show statistically significant differences between all six pairs of criteria (all $p~<~0.001$). 

Interview scripts show that while participants ranked reusability as the least important factor, participants were reusing PINs for reasons of memorability.
\begin{changemargin}{0.2cm}{0.2cm} 
\textit{``Its really annoying to have to remember a new PIN so I change them all to the one I was using. I wouldn’t be able to keep track of what PIN is for which card if I didn’t make them all the same. I have five cards that have PINs.''} (P28)
\end{changemargin}

\begin{table}
  \caption{Reported reasons for 49 PIN updates}
  \label{tab:pin_update}
  \centering
  \small
  \begin{tabular}{l | c c c}
    {\small\textit{}}
    & {\small \textit{Digital}}
    & {\small \textit{Financial}}
      & {\small \textit{Physical}} \\
    {\small\textit{Reason for update}}
    & {\small \textit{n=22}}
    & {\small \textit{n=18}}
      & {\small \textit{n=9}} \\
   \hline
    Security (preventive) & 5 & 7 & 3 \\
    Security (post-compromise) & 3 & 5 & 1 \\
    Easy to remember & 1 & 4 & 2\\
    Forgot the PIN & 3 & 1 & 1\\
    Policy requirement & 2 & 0 & 0\\
    Impulse & 1 & 1 & 0\\
    Asset upgrade & 7 & -- & 2\\
  \end{tabular}

\end{table}

\subsection{PIN Update}
For each asset category, we asked participants to recall the last time they updated a PIN. 
Participants reported 49 incidents of PIN changes (22, 18, and 9 for digital, financial, and physical assets, respectively).
The number of reported PIN updates differed significantly between asset groups 
($\chi^2(2) = 6.58, p~<~0.05$).

Table~\ref{tab:pin_update} shows that 9/49 (18\%)  PIN updates across asset types were due to the compromise of PINs.  
In Section~\ref{subsec:pin_attacks}, we report our findings that the majority of PIN compromises do not result in a PIN update.
Another 15/49 (31\%) PIN updates were performed as a preventive security measure.
The reasons for the update were similar to the following: \looseness=-1
\begin{changemargin}{0.2cm}{0.2cm} 
\textit{``Yes, changed it because felt it was good to change. Because its more secure to change it from time to time.''} (P16)
\end{changemargin}
Five of the seven participants who updated PINs of their financial asset did so to change the default PIN that was set by the bank for security reasons.
 
12/49 (24\%) of PIN updates were due to memorability reasons---either motivated by participants' decision to choose easy to remember PINs or as a result of forgetting a PIN.
\begin{changemargin}{0.2cm}{0.2cm} 
\textit{``I have been using this PIN for various things for 30 years. [I] set my devices to the same PIN when I get them.''} (P31)
\end{changemargin}
For digital assets, 7/22 (32\%)  PIN updates were a result of a device (smartphone) upgrade.
The reason for PIN updates due to device upgrades are explored below.
Other less common reasons included policy requirements and impulsive updates.

\begin{figure}[t]
  \centering
  \includegraphics[width=.8\columnwidth, trim={15mm 5mm 15mm 5mm},clip]{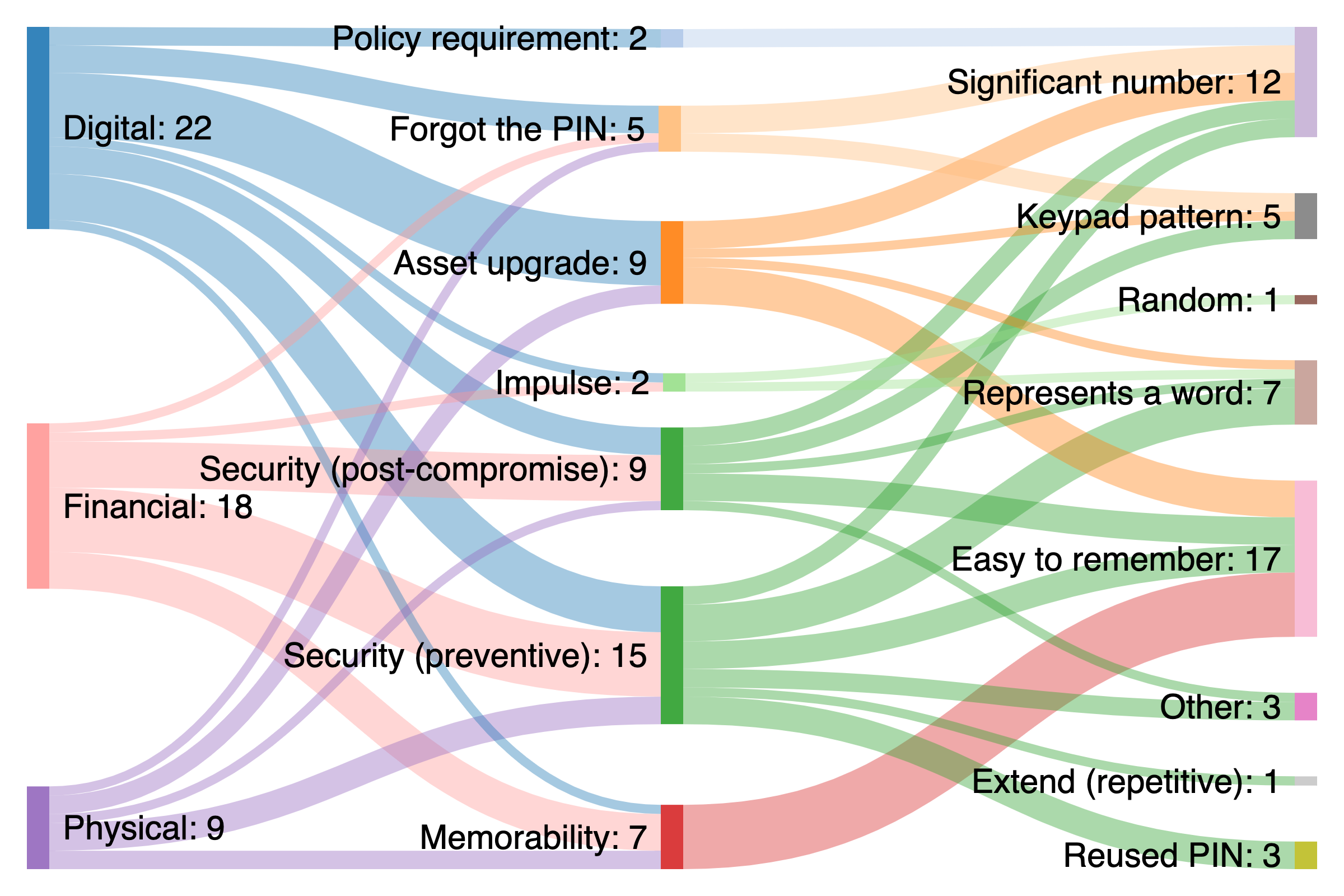}
  \caption{Participants' reasons for PIN update (labels in the center) and strategies for PIN updates (labels on the right).}
  \label{fig:pin_update_strategies}
\end{figure}

Participants were asked to describe the strategy they used to pick the new PIN for each of the PIN update events, and we furthered queried participants about different events that led to PIN updates.
The codified responses are reported in Figure~\ref{fig:pin_update_strategies}, which  shows that the most popular strategies for selecting a new PIN are choosing an easy to remember number or a significant number, such as a date.
Other popular strategies included using numbers that represented a word or using a pattern on the keypad. 17/49 (35\%) PIN update strategies were simply reported as an easy to remember PIN. 
Since patterns or reused PINs are easy to remember, it is not clear how many of these participants were choosing patterns or reusing other PINs.
We discovered this confound during our analysis; therefore, for digital assets, we were unable to collect data on what prompted participants to update PINs when they acquired a new device. 
However, the update strategies show that users employed approaches that result in better memorability (easy to remember or represents a significant number or word).
The two cases for physical asset upgrades are reported for situations when participants moved to a new place and updated PINs for digital locks.

It is interesting to note that significant numbers and keypad patterns were popular PIN update strategies despite the reason for update---whether it was security or memorability. 
Insecure PIN selection strategies were prevalent in high-risk scenarios:
\begin{changemargin}{0.2cm}{0.2cm} 
\textit{``Yes, suspect my ex-girlfriend had it. I think she saw me enter it in and I changed it after that. I added two digits to the old one to make it a six digit PIN. I added a repeat two digits to the end of the PIN''} (P16)
\end{changemargin}

\subsection{PIN Sharing}

Two factors that possibly influence PIN sharing habits include the type of asset (e.g., home lock vs. smartphone PIN) and co-habitation. 
For the former, we separately report the sharing habits for different asset types.
For the latter, we asked participants whom they lived with: seventeen participants reported living with a spouse, seven with roommates, four with a romantic partner, three with parents, three with siblings, twelve with children, and two by themselves. \looseness=-1

The reported statistics for PIN sharing are provided in Table~\ref{tab:pin_sharing}. Only a few participants reported not sharing PINs with anyone.
For digital assets, only 6/32 (19\%)  participants did not share their PINs.
21/32 (66\%) participants shared PINs for their digital assets with other people that they were in a romantic relationship with. 9 (28\%), 5 (16\%), and 5 (16\%) of the 32 participants shared their digital PINs with children, friends, and siblings, respectively.
Four participants reported sharing PINs of their digital assets because they were in circumstances where they felt that they had no other option but to share it temporarily. However, all of them reported not updating PIN after sharing for trust or other reasons:
\begin{changemargin}{0.2cm}{0.2cm} 
\textit{``I had to share it with my step-child once that I was driving. I thought about changing it but not too keen on changing it since new PINs are a hassle. He visits us once a week only so that is also a factor.''} (P26)
\end{changemargin}

\begin{table}[t]
\caption{People that participants reported sharing PINs with.}~\label{tab:pin_sharing}
  \centering
  \small
  \begin{tabular}{l | c c c}
    &  \textit{Digital} &  \textit{Financial} & \textit{Physical} \\
    \textit{Shared with} &  \textit{(n=32)} & \textit{(n=34)} & \textit{(n=27)} \\ \hline
    None & 6 & 7 & 1 \\
    Spouse & 16 & 17 & 13 \\
    Children & 9 & 6 & 8\\
    Parents & 3 & 7 & 10\\
    Siblings & 5 & 2 & 5\\
    Girl/Boyfriend & 5 & 3 & 4\\
    Friends & 5 & 2 & 9\\
    Helpers & 0 & 0 & 8\\
  \end{tabular}
  \vspace{-.1in}
\end{table}

For financial assets, 7/34 (21\%) participants reported not sharing their PINs with anyone. 
Participants mostly shared their financial asset PINs with their romantic partners (20/34 (59\%)) and parents (3/34 (21\%)).
Only two participants reported sharing with friends 
to grab lunch or coffee for them.
Similar to digital PINs, three participants reported inadvertent sharing of financial PINs and not updating them later.
\begin{changemargin}{0.2cm}{0.2cm} 
\textit{``I have given it to my son once too to buy something and was concerned if he would try the same on my laptop or smartphone. Yes, he knows those PINs now [laptop and smartphone PINs---the same as their ATM card] but back then he didn’t. Was holding another kid and there was an urgent need to grab water from convenience store.''} (P24)
\end{changemargin}

As expected, for physical assets, all but one participant shared their PINs with at least one other party. Other than prevalent sharing among friends and family members, 8/27 (30\%) participants reported sharing physical PINs with hired helpers (cleaners or pet caretakers).
For physical PINs, two cases of inadvertent sharing with strangers were identified.  Participants reported not updating the PIN, even after their contact with the third parties had concluded. We discuss the reasons for not updating physical PINs in Section~\ref{sec:discussion}.
\begin{changemargin}{0.2cm}{0.2cm} 
\textit{``Yes, it was shared with the furniture company that went out of business, and nothing was done about it. Never had a problem with the company so there is a trust.''} (P11)
\end{changemargin}

The self-reported sharing data on PINs from our study shows widespread sharing as well as sharing across different relationship types.
Among participants that live with their spouses, all but one (94\%) shared their PINs for digital assets. For participants who reported living with a girlfriend/boyfriend, all shared their digital PINs with their partner.
This finding is different from Kaye's finding that only a third or fewer  participants reported sharing their personal email and Facebook passwords, both primarily with partners and close friends~\cite{kaye2011self}. 
This difference is expected because Kaye studied sharing habits for specific online services whereas, participants from our study reported sharing habits for assets that are either more likely to be shared (e.g., physical) or assets that are more generic in nature (e.g., smartphones). 
Our findings are congruent with those of Matthew et al.~\cite{matthews2016she} and Singh et al.~\cite{singh2007password} that people share passwords with trusted family members.

\begin{table}
\caption{Reported reuse of PINs}
\label{tab:pin_reuse}
\centering
\small
    \begin{tabular}{c c l}
\multicolumn{3}{c}{\textit{Have you reused PINs?}}    \\ \hline \hline

\multicolumn{2}{r}{\textbf{No:}}& 7/35 (20\%)\\
    \multicolumn{2}{r}{\textbf{Yes:}}& 28/35 (80\%) \\
    \hline
    \multicolumn{3}{c}{\textbf{Type of reuse}}  \\
\multicolumn{3}{c}{18/28 across all asset types} \\
\multicolumn{3}{c}{3/28 same asset type only} \\
\multicolumn{3}{c}{4/28 across digital and physical } \\
\multicolumn{3}{c}{3/28 across digital and financial } \\
\hline\hline
\end{tabular}
\vspace{-.2in}
\end{table}

\subsection{PIN Reuse}
\label{subsec:pin_reuse}
We asked participants whether they reuse PINs (within the same or across asset categories). 
Seven (20\%) participants reported not reusing PINs at all for security reasons.
These seven participants used 1, 2, 3, 5, 5, 6, and 6 PINs in total. 

Twenty-eight participants reported reusing PINs.
As discussed in Section~\ref{subsec:pin_choices}, the reported underlying reason for reuse was memorability.
Out of these participants, 18 (64\%) reported reusing PINs across all categories.
Three participants reported reusing PINs within the same category only (e.g., a common PIN for both their phone and tablet).  Findings also showed that three participants reported reusing PINs across digital and physical categories, and the same number reported reusing across digital and financial asset categories.  Participants' choice to not reuse PINs and create new PINs for some assets was motivated to protect against certain threats. \looseness=-1
\begin{changemargin}{0.2cm}{0.2cm} 
\textit{``[I have] shared PIN-A [(Cell phone, laptop, online account (cell provider))] and PIN-C [(Netflix parental, xBox)] with spouse and PIN-B with kids [(Home, Garage, tab)].''} (P25)
\end{changemargin}

During the interviews, several participants demonstrated that they understood the risk of reusing PINs. However, they either considered the reuse to be a secret or a chance worth taking despite the risks involved.

\begin{changemargin}{0.2cm}{0.2cm} 
\textit{``Well it is the same as my garage and alarm PIN. All financial PINs are the same so I have shared it with my wife, kids, dog walker, and cleaning lady but only my wife knows it’s the same PIN for my bank.''} (P28)
\end{changemargin}

\begin{figure}[t]
    \centering
    \includegraphics[width=64mm, trim={0 5mm 0 2mm},clip]{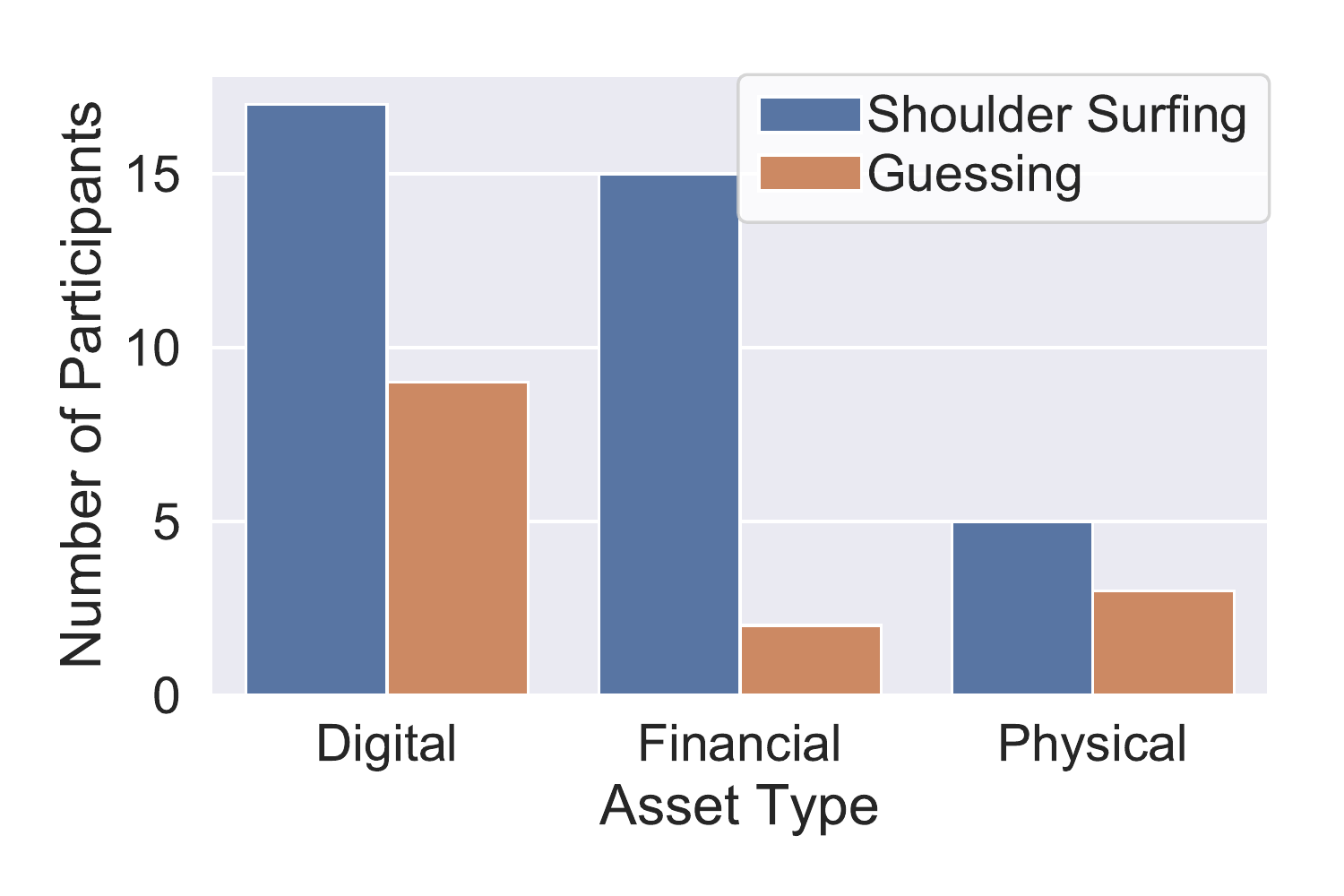}
    \caption{Participants reported attacks on PINs for different asset categories.}
    \label{fig:pin_attacks}
  \end{figure}

\begin{figure}[t]
  \centering
  \includegraphics[width=0.9\linewidth, trim={15mm 5mm 15mm 5mm},clip]{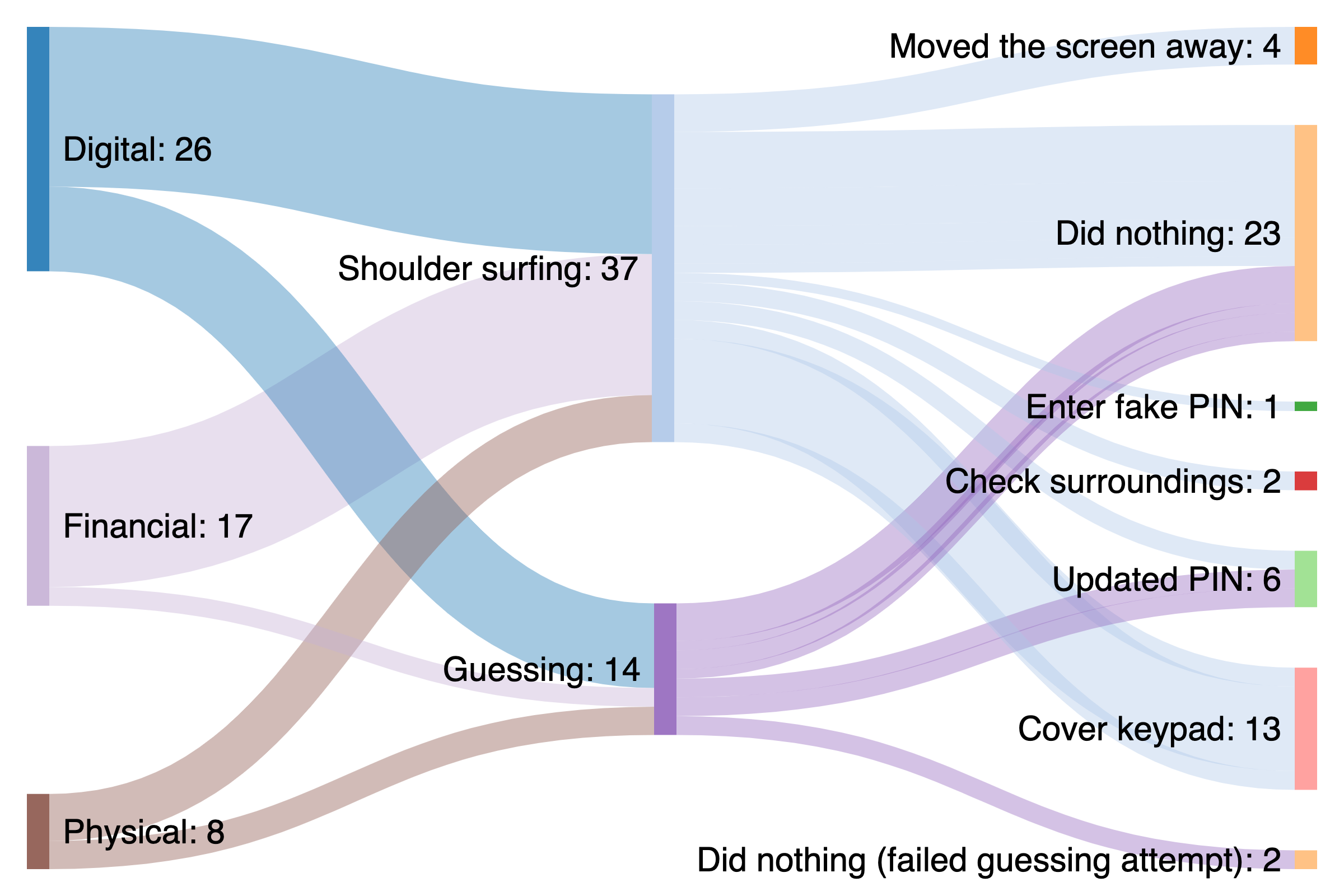}
  \caption{Reported recourse by participants (labels on right) against different attacks (labels in center) on PINs.}
  \label{fig:pin_recourse}
  \vspace{-.1in}
\end{figure}


\subsection{Attacks on PIN and Recourse}
\label{subsec:pin_attacks}
We asked participants to recall the last time a PIN in each asset category may have been subjected to shoulder surfing or guessing attacks, regardless of the outcome of the attacks.
Figure~\ref{fig:pin_attacks} shows the distribution of participants who perceive that an attack may have occurred on their PINs of different categories. 
25 participants (71\%) recall experiencing a shoulder surfing attack or being concerned that a guessing attack had occurred on one or more of their PINs. 
More participants reported attacks on PINs for digital and financial assets than for physical assets (26 and 17, respectively vs. 8).
Similarly, across all asset types, participants reported more shoulder surfing attacks than guessing attacks (37 vs. 14). 
The number of shoulder surfing and guessing attacks reported both differed significantly between asset types ($\chi^2(2) = 7.74, p~<~0.05$ and $\chi^2(2) = 6.84, p~<~0.05$, respectively). 
Significantly more shoulder surfing attacks were reported for financial assets than guessing attacks ($\chi^2(1) = 11.29, p~<~0.05$). 




We also asked participants regarding the recourse that they took when they were subjected to  attacks.
Their responses were codified, and a summary of results is presented in Figure~\ref{fig:pin_recourse}.
For 23 of the potentially successful attacks, participants reported taking no action to prevent shoulder surfing or guessing attacks across different asset categories.
The reasons reported for this inaction included trusting the attacker (friend or family member), laziness, or that the attack failed, so they felt action was not required. 
\begin{changemargin}{0.2cm}{0.2cm} 
\textit{``Yes, at work some colleagues have seen me enter it. Usually for meetings I have to open my device and enter PIN in front of other people. I am not looking but can imagine that every one who is next to me have seen it. Did not change it and did not protect because it seems like people would think that I do not trust them.''} (P35)
\end{changemargin}


Six participants reported updating PINs in response to an attack---four for digital assets and two for financial assets. 
Note that when participants were asked about why they updated a PIN (for a previous question, see Table~\ref{tab:pin_update}), more participants reported updating the PIN due to reasons of security.
However, that over-reporting is due to compromises through other types of attacks (e.g., online compromise of a PIN-protected financial asset).
Other common defences included covering the device screen or moving the screen away from the attacker (similar to the finding of Eiband et al.~\cite{eiband2017understanding}).

%% file: 05-discussion.tex





\section{Discussion}
\label{sec:discussion}

Our interviews uncovered interesting ways in which different asset categories can impact PIN management and 
unique security and memorability challenges for PINs. 
In this section, we discuss these issues. 
For qualitative analysis, two researchers independently performed thematic analysis to identify themes from participant responses during the semi-structured interview.  Identified themes were compared and discussed by reviewers until consensus was reached. This approach is used by other researchers in the field (e.g., Acar et al.~\cite{acar2017internet}).


\subsection{PINs in Different Contexts}
Our findings show differences between different asset types in how participants share PINs and how PINs are attacked.
Participants also reported using different levels of protection for different types of assets.
This behaviour was due to different levels of perceived risk to different types of assets and the possible recourse available to the users in case the attacker was successful.
For instance, five participants reported being less concerned about physical PINs than digital or financial PINs and had comments similar to the following.
\begin{changemargin}{0.2cm}{0.2cm} 
\textit{``Even if it was access to where the digital devices or money [financial] PINs are [through physical PINs], the risk of breaking a physical PIN is higher for getting caught then the other ones.''} (P16)
\end{changemargin}
Similar comments were from two participants who were less concerned about other people learning their financial PINs.

\begin{changemargin}{0.2cm}{0.2cm} 
\textit{``I would be a lot more concerned about someone accessing my phone than my bank account. If a colleague were to look into my phone or laptop I would not have a recourse but if someone were to steal my money that will be a different thing.''} (P35)
\end{changemargin}
Two participants also reported caring less about their financial PINs because an attacker with their bank cards will be able to perform transactions without needing the PIN.

\begin{changemargin}{0.2cm}{0.2cm} 
\textit{``Not much [worried about PIN security] and I guess it is a combination of factors. [Bank] card is on me and if someone were to get it they could tap-to-pay or do an online transaction with the number on the back. And in that case there is reimbursement for fraud.''} (P35)
\end{changemargin}

Only one participant reported being more concerned about the physical PINs due to their perceived susceptibility to shoulder surfing attacks.
\begin{changemargin}{0.2cm}{0.2cm} 
\textit{``Yes. I am worried someone would watch me enter it. They may have binoculars. I always cover it [hand masking entry].''} (P29)
\end{changemargin}


The comments of participants indicate that with the availability of possible recourse (i.e., police  involvement for physical or financial assets), they were less careful about the secrecy of their PINs. 
Egelman et al.~\cite{egelman2014you} also reported observing this rational behaviour for the use of security features on smartphones and risk perceptions of users.
However, while the perception that the attackers are less inclined to trespass on their property may be true, the majority of participants reported reusing their PINs across other categories.

\subsection{Attack Susceptibility of PINs}
In Section~\ref{subsec:pin_attacks}, we reported our finding that 27/35 (71\%) of participants reported attacks on PINs.
Another interesting theme that emerged from participants' responses was the high susceptibility of PINs to shoulder surfing attacks.
Three participants voiced the concern that it is difficult to enter a PIN without third parties in close proximity learning about it.

\begin{changemargin}{0.2cm}{0.2cm} 
\textit{``My colleagues may know my PIN but not too sure whether it is worth changing it because they will learn the new one too. Mostly this happens when you unintentionally look at someone entering it.''} (P24)
\end{changemargin}
This observation was also the reason why two of the participants did not update their PINs after they were compromised.
\begin{changemargin}{0.2cm}{0.2cm} 
\textit{``My kids are not supposed to know it but they must have seen me enter it on my previous phone when I did not have a fingerprint id. [...] I am not too sure who else has knows it or has cared to learn it. I have seen many enter their PINs and patterns before me but never cared for it.''} (P33) \looseness=-1
\end{changemargin}

During the discussion on guessing attacks, the comments of four participants seemed to indicate that they understood that their PINs were weak and could be easily guessed.

\begin{changemargin}{0.2cm}{0.2cm} 
\textit{``Didn’t ask how he [the perpetrator] got to know but I guess he watched me type it or he may have guessed it since it was simple enough.''} (P25)
\end{changemargin}


Participants' comments show that they relied on other measures to complement security offered by PINs. These approaches include risk aversion of attackers against attacking financial and physical assets (discussed earlier), aversion of attackers to be recorded in the act, and participants being careful of their assets around attackers: 


\begin{changemargin}{0.2cm}{0.2cm} 
\textit{``For the gym locker PIN, I am worried sometimes because many people are around and I leave my wallet and phone in the bag when going for shower. But there are cameras in some areas so I think people would not try something silly.''} (P33)
\end{changemargin}


Two participants complained about the PIN entry interface for Netflix Parental lock. 
These participants complained that on big screens, the Netflix parental lock did not provide them with a way to enter the PIN without giving it away in shared spaces---particularly with the children in the vicinity. 
\begin{changemargin}{0.2cm}{0.2cm} 
\textit{``[...] when my kids ask me to play specific content  I've to ask them to leave the room.''} (P24)
\end{changemargin} 

The relative ease with which PINs can be shoulder surfed is known~\cite{khan2018evaluating}. Our study shows users are aware of this issue, and that it negatively affects trust in PINs as an effective security control.
We discuss some remediation in Section~\ref{sec:guidelines}.

\subsection{Memorability Issues}
As noted, participants rated memorability as the most important criteria for selecting PINs. This high ranking may be attributed to avoiding potential inconveniences:

\begin{changemargin}{0.2cm}{0.2cm} 
\textit{``You have to be really quick in restaurants or stores, you can’t be guessing and trying to remember it. That’s why I keep the same PIN.''} (P32)
\end{changemargin}

Memorability and ease of entering a known PIN seemed to trump security even for the cases where participants decided to update PINs. Three participants reported that they reluctantly reverted their PINs because of frequent errors. 

\begin{changemargin}{0.2cm}{0.2cm} 
\textit{``Did change after [my girlfriend learned it] because we were living in the same shared space but made so many mistakes that I reverted; entry mistakes from muscle memory''} (P27) \looseness=-1
\end{changemargin}


Stobert and Biddle~\cite{stobert2014password} found that users found coping mechanisms to live with the difficulties of password authentication. Similarly, PIN users seem to be using strategies to deal with the memorability-related challenges of PIN authentication by compromising security.
In Section~\ref{sec:guidelines}, we discuss some approaches to mitigate these memorability-related challenges.

\subsection{PINs and Past Relationships}
Park et al.~\cite{park2018share} conducted an online survey and found that, among other factors, marriage and co-habitation results in the sharing of online accounts. 
Our findings are congruent with theirs.  
Our participants self-reported wide-spread sharing of PINs with their romantic partners.

Nine participants reported sharing their digital PINs with someone that they were in a romantic relationship with in the past. 
Three of these participants did not change the PIN because they either still trusted that person or they felt there was no need since the other person no longer had access to assets (\textit{``[I] changed it just for more privacy but didn’t feel the need to change it.''} (P6)).
Other participants updated their PINs, although one participant reported that there was no need to do so (\textit{``[Did] nothing as I had the device}'' (P3)).
One participant reported changing PIN because the other person still cohabited with them.

For financial assets, five participants reported sharing it with people that they were in a relationship in the past, and only two people reported updating it.
Note that these participants also reported updating their digital PINs after moving on.
Only four participants shared their physical PINs with past relationships, and only one reported updating it.
While these PINs were for home or garage access, participants reported not changing those because they still trusted their past partner (\textit{Nothing was done as there was never a problem.} (P13)).

Park et al.~\cite{park2018share} identified that individuals are likely to attempt to remove or disable a partner's access to online accounts. 
We did not find this to be the case for our participants.
Unlike with online accounts, participants would need access to assets in addition to the authentication secret (i.e., PINs).
However, with the increasing number of online services that accept PINs and the widespread reuse, this may pose a threat to those accounts where PIN has been reused. 
For such cases, it would be beneficial to consider the guidelines suggested by Obada-Obieh et al.~\cite{obadaburden2020burden} on design improvements of online accounts to support users better when they end account sharing.\looseness=-1 


\subsection{Physical PIN Inheritance and Update}
One interesting finding was the ``inheritance'' of physical PINs that protected garage doors. 
Nine participants reported moving to another house with a pre-existing PIN set to open the garage door, but only three participants reported updating that PIN while the remaining six kept the PIN set by the previous owner. 
One participant even reported reusing the inherited PIN for their home lock: \looseness=-1
\begin{changemargin}{0.2cm}{0.2cm} 
\textit{``Garage [PIN was set] by previous owner. [I] used it again for home lock that was installed afterwards''} (P27)
\end{changemargin} 
Since four of these six participants reported changing the home locks, the lack of the update of garage door PINs cannot be attributed to trust.
Instead, this insecure behaviour is due to the lack of knowledge on how to update the PIN:
\begin{changemargin}{0.2cm}{0.2cm} 
\textit{``No, the garage was setup by previous owner. We did change the key locks and considered updating the garage PIN but there is no information available on it on how to do that.''} (P34)
\end{changemargin} 
This inability to update garage door PIN was also voiced by participants when these PINs were accidentally divulged:
\begin{changemargin}{0.2cm}{0.2cm} 
\textit{``Once a person who was delivering a package [saw it]. My husband was concerned about it but neither knew how to change it.''} (P35)
\end{changemargin} 
While the instructions on how to update these PINs were missing, two participants did comment that laziness on their part also contributed to the situation, and that they had other resources available.
\begin{changemargin}{0.2cm}{0.2cm} 
\textit{``[It was shared with the] Garage door repair person when they were here to fix the door. Didn’t change it... don’t know how to although I can google [search].''} (P26)
\end{changemargin} 
One participant complained that the previous owner did not share the Master PIN that would allow a PIN update, thereby eliminating their ability to update it.
The inability of users to effortlessly update PIN in case of a compromise could potentially result in security issues.
In Section~\ref{sec:guidelines}, we discuss possible remediation strategies.  

%% file: 06-future_work.tex
\section{Future Research Directions}
\label{sec:guidelines}

\paragraph{PIN choices and management strategies.}
Our participants reported widely sharing and reusing PINs, and infrequently changing them  even after they were compromised.
The interviews indicate that the main driving factor behind this risky behaviour was the memorability of PINs.
Most participants did not adopt a PIN management strategy by explicitly considering the threat actors.
When prompted to choose a PIN, they chose a PIN that they remembered well.
Only a few participants considered aspects such as the circles they had to share the PIN with before choosing their PINs.
Other factors that need to be considered include the nature of the asset, the susceptibility of attacks on the asset (e.g., shoulder surfing is more of a threat for a smartphone than an ATM PIN), and the type of recourse that is available to participants in the event of a compromise.
While these are important considerations, additional research needs to be conducted to understand that a user with an average technology and security proficiency is able to make secure PIN choices given these factors.
This will enable researchers to create improvements that actually match user expectations in their everyday lives.

PIN-based authentication is used for six assets on average and recalling the correct PIN for the right asset is problematic for several participants.
Existing proposals on the memorability of PINs (discussed in Section~\ref{sec:related_work}) do not improve the situation with multiple assets and multiple PINs.
A cued recall-based approach that allows a participant to associate pairs of assets and PINs (or corresponding word representation of PINs) may offer mitigation. 
Digital wallets, for example, enable users to perform secure transactions without entering PINs, but such features are not available for all PINs, particularly physical PINs.
Digital apps for smartphones could be designed to help people with such recall issues with features similar to that of a password manager but would enable quick recall for digital, financial, and physical PINs. 

\paragraph{Confidence in PINs as a security control.}
Prior empirical studies report on the susceptibility of PINs to shoulder surfing attacks and users' experiences of such attacks~\cite{aviv2017towards, eiband2017understanding,harbach2014sa, khan2018evaluating}. 
We also uncovered the limited levels of faith users reported on PINs' resistance to shoulder surfing attacks (see Section~\ref{subsec:pin_attacks}).
While simple defences like shielding the keypad while entering a PIN is effective, it is not widely used as it shows the lack of trust to the observers.
Improved PIN entry interfaces have been proposed that provide defences against shoulder surfing (discussed in Section~\ref{sec:related_work}), but the limited availability of these on smartphones may reduce the efficacy of PINs as an effective security control. 
We also noted several cases where participants had to inadvertently share their PINs or enter PINs in front of other people.
The availability of a short-term device access approach like SnapApp~\cite{buschek2016snapapp} may help users greatly improve the security posture of their digital PINs.

\paragraph{Improved interfaces for PIN update.}
PIN-based authentication on devices with limited interfaces (e.g., garage doors and digital home locks) introduces unique challenges. 
Our study shows that users are more likely to continue inherited PINs for such assets due to the lack of clear and readily available instructions on how to update PINs.
Furthermore, such assets may require a master PIN to update or reset PINs, and the storage and management of such a PIN further complicates the situation.
One participant reported sharing the same PIN with people of different trust levels with (e.g., family vs. pet caretaker) despite the availability of the digital home lock to create different PINs.
This was primarily due to the inability of the device to report which PINs were used when.

As the security of an asset is dependent on being able to change the PIN in case of a compromise, there is a need to design a standard way to update and reset PINs on devices with limited interfaces (i.e., only keypad).
Alternatively, instructions could be provided on the physical locks to reduce barriers to PIN update on such devices. 
While the availability of such unifying methods for updating PINs on future devices would make a difference in encouraging PIN updates after compromise, the challenge will remain for millions of devices currently in use. 
One possible approach is to design augmented reality tools to  address this gap by linking these physical assists to known documentation and instructions for updating PINs.


%% file: 07-limitations.tex
\section{Limitations}
\label{sec:limitations}

Our study has some inherent limitations similar to that of other user studies, which include that many of the findings are based on self-reported data from willing participants. Prior empirical studies of PIN usage on smartphones~\cite{harbach2014sa} indicated that participants under-report their daily PIN usage, which may also be the same here. In which case, our results may underestimate the total number of PINs used across asset types, which is compounded by the fact that some categories, such as banking app PINs, could be classified as both digital and financial. In an attempt to mitigate this limitation, we choose to use a semi-structured interview method that included suggestions of assets, to help ensure that participants thought of the diversity of assets where PINs are used. 


Additionally, we asked several contextual questions for the most widely used PIN in each category. As a result, our study is limited in scope with regard to the most widely used PINs, but we were able to collect quality responses from participants in a time-constrained lab-based study regarding the PINs that protect the most assets. 

We were also limited geographically in our participant pool, which belonged to the Waterloo and Guelph regions in Canada. This is a relatively safe place to work and live (as self-reported by the participants). The safe environment may have implicitly encouraged some of the unsafe practices among our participants for PINs protecting their physical assets. However, we do believe this convenience sample does generalize to many other populations, but not all, and more research would be needed to understand how different populations approach PIN security.

Finally, interviews were conducted by two researchers sequentially, where both researchers used the same semi-structured script of questions. We found that the second researcher elicited more detailed quotes from participants, which are cited more throughout the document; however, data collected by the first researcher are still ecologically valid and were fully used in data analysis.

%% file: 08-conclusion.tex
\section{Conclusion}
\label{sec:conclusion}
We conducted a study with 35 participants to understand how they manage PINs across different assets.
Our findings show behaviour that may result in potential compromises due to widespread sharing and reuse of PINs across different asset categories were mainly motivated by reasons of memorability. The memorability concerns also deter users from updating PINs after they are compromised.
Participants further reported their lack of confidence in PINs due to their susceptibility to shoulder surfing attacks---a concern that can be mitigated using PIN entry interfaces that resist shoulder surfing.
Our study also shows that participants change their PIN management behaviour for different types of assets due to the availability of another recourse in case of a compromise. 
Finally, we propose further research directions for researchers. 
With the increasing options to use PINs for purposes of authentication for different types of assets, our findings will help researchers design tools and strategies to improve the security of PIN-protected assets.

%% file: 99-appendix.tex
\section*{Appendix}
\label{sec:appendix}

\section{Survey Material}

\subsection{Closed Response Demographic Questions}
\label{app:demo}

\begin{enumerate}
	\item What is your age?
	
(a) 18-25; (b) 26-30; (c) 31-35; (d) 36-40; (e) 41-45; (f)~46-50; (g) 50+ yrs; (h) Prefer not to answer

	\item What is your identified gender?
	
(a) Male; (b) Female; (c) Non-binary; (d) Other; (e) Prefer not to answer

	\item What is your highest level of education?
	
(a) Some high school; (b) High school; (c) Some college/university; (d) Trade/technical/vocational training; (e) Associate’s degree; (f) Bachelor’s degree; (g) Master’s degree; (h) Professional degree; (i) Doctorate; (j)~Prefer not to say

	\item What is your annual household income?
	
(a) Under \$15,000; (b) \$15,000 -- \$29,000; (c) \$30,000 -- \$49,999; (d) \$50,000 -- \$74,999; (e) \$75,000 -- \$99,999; (f) \$100,000 -- \$150,000; (g) over \$ 150,000; (h) prefer not to answer

	\item Which of the following best describes your educational background or job field? 
	
	(a) I have an education in, or work in, the field of computer science, computer engineering or IT; \\(b) I do not have an education in, nor do I work in, the field of computer science, computer engineering or IT; \\(c) prefer not to answer
	
	\item Which of the following best describes your level of proficiency with technology?
	
(a) Basic (I can perform basic tasks on a smartphone/laptop such as sending emails or browsing the internet; \\
(b) Intermediate (I can perform intermediate tasks on a smartphone/laptop such as changing the settings or installing new applications); \\
(c) Advanced (I have knowledge of and am capable of writing source code); \\
(d) Prefer not to answer

	\item Which of the following best describes your level of proficiency with security?
	
(a) Basic (I have a limited understanding of security i.e., does not know what antivirus is or does not know how to use it); \\
(b) Intermediate (I have some knowledge on aspects of security and different threats that exist and how to remediate some of them); \\
(c) Advanced (I have some formal training or actively researches security topics); \\
(d) Prefer not to answer

	\item Can you identify which of these relationships types apply to those you currently live with (choose all that apply)?
	
(a) Alone; (b) Spouse; (c) Own children; (d) Parents; (e) Siblings; (f) Friends; (g) Roommates; (h) Other (please describe the relationship type); (i) Prefer not to answer

	\item I live in an area that is: (\emph{5-point Likert scale ``Very safe''-- ``Very unsafe''})

	\item I work or spend time in an area that is: (\emph{5-point Likert scale ``Very safe''-- ``Very unsafe''})

\end{enumerate}

\subsection{Semi-structured Interview Questions}
\label{app:semi}
\subsubsection*{Asset Category Independent Questions: Part I}

\begin{enumerate}
	\item Can you please tell us how many unique PINs you currently use?

	\item Which resources do those PINs protect? For ATM cards, does the PIN protect one or multiple ATM cards?

    [Participants were reminded of some assets that are commonly protected by a PIN. The list included ATM cards, smartphones, laptops, personal computers, online accounts, electronic home locks, home security systems, garage door openers, cars, bike/gym locks, voicemail, gaming consoles, apps, smartwatches, thermostats, and other home devices.]

	\item Did you miss any PINs previously? What resources do they protect?

\end{enumerate}
\subsubsection*{Asset Category Dependent Questions}
The following questions were repeated for each of the three PIN categories (digital, financial, and physical). Context-dependent questions were for the most frequently used PIN in each category, unless otherwise noted.

\begin{enumerate}
	\item Who else have you shared this PIN with? If friends or roommates, how many?\par

	\item How concerned would you be if your PIN was revealed to the following people: (\emph{5-point Likert scale ``Very concerned''-- ``Not concerned at all''}) 
	
	(a) Friends, (b) Roommates, (c) Parents, (d) Siblings, (e) Spouses, (f) Children

	\item How long have you been using this PIN?

	\item Have you ever changed a PIN in this category in the past? If so, what prompted it? 
	
	\item \textbf{[IF CHANGED PIN]} What are your strategies for changing a PIN and picking a new PIN?\par

	\item Can you rank how important the security of this asset is you? (\emph{5-point Likert scale ``Very important''-- ``Not important at all''})

	\item When picking the PIN, what was the order of importance for the following criteria: (a) memorability; (b) ease of usability; (c) security; (d)  reuse of a previous PIN

	\item How often do you enter a PIN for this category? 

	\item (For any asset in this category) Has there ever been a situation where someone learned your PIN? If so, who? How did they learn it? What device? What was your recourse? If the PIN was not updated, why?

	\item (For any asset in this category) Have you ever been in a situation where you were worried about someone observing your PIN? What was your recourse?

	\item (For any asset in this category) Have you ever been in a situation where you were worried about someone may try guessing your PIN? What was your recourse?

	\item (For any asset in this category) Have you ever shared a PIN with someone in the past that you are no longer in a relationship with? If so, who? [Examples include past spouses, friends, coworkers, and roommates]

	\item \textbf{[IF SHARED WITH PAST RELATIONSHIPS]} Did you take any steps to ensure that such people no longer have access to your PIN protected resources? What steps did you take, and why?

	\item Have you ever tried to learn or observe a PIN of someone? How? Was it successful? What resource were you trying to access? Was it someone you knew?

	\item If we ask you to guess the PIN of a person you know in five guesses, what strategies will you take?

	\item Would your strategies for the above question change if it was a stranger?

	\item (For any asset in this category) Do you store or write PINs anywhere, like a notepad or online password manager? If you write them on a notepad, where do you store it?
\end{enumerate}

\subsubsection*{Asset Category Independent Questions: Part II}

\begin{enumerate}
    \item Have you ever used the same PIN for two or more devices? How about devices that are in different classes?\\
    
    \item Consider a digital device you use to login to your banking website/app or your digital wallets like Apple Pay or Google Pay. Is this device protected using a PIN (including PIN backup for fingerprint? If so, does access to your banking website or your digital wallets require another PIN or a password?
      
    \item How were the physical PINs to home or garage access were setup? 
     [Did they set them up? Did they updated them when they moved to a new place? Did a technician set them up? Did the previous owner set them up or is it the default PIN? If the previous owner set up the garage door PIN, did they change key locks? If so, why not other PINs?]\\
\end{enumerate}


%% file: main.bbl
\begin{thebibliography}{10}

\bibitem{abdelrahman2017stay}
Yomna Abdelrahman, Mohamed Khamis, Stefan Schneegass, and Florian Alt.
\newblock Stay cool! understanding thermal attacks on mobile-based user
  authentication.
\newblock In {\em Proceedings of the 2017 CHI Conference on Human Factors in
  Computing Systems}. ACM, 2017.

\bibitem{acar2017internet}
Yasemin Acar, Michael Backes, Sascha Fahl, Doowon Kim, Michelle~L Mazurek, and
  Christian Stransky.
\newblock How internet resources might be helping you develop faster but less
  securely.
\newblock {\em IEEE Security \& Privacy}, 15(2):50--60, 2017.

\bibitem{amitay2011common}
{Amitay, Daniel}.
\newblock {Most common iPhone passcodes.}
\newblock
  \url{http://danielamitay.com/blog/2011/6/13/most-common-iphone-passcodes},
  2011.
\newblock Last accessed June, 2020.

\bibitem{aviv2017towards}
Adam~J Aviv, John~T Davin, Flynn Wolf, and Ravi Kuber.
\newblock Towards baselines for shoulder surfing on mobile authentication.
\newblock In {\em Proceedings of the 33rd Annual Computer Security Applications
  Conference}. ACM, 2017.

\bibitem{biddle2012graphical}
Robert Biddle, Sonia Chiasson, and Paul~C Van~Oorschot.
\newblock Graphical passwords: Learning from the first twelve years.
\newblock {\em ACM Computing Surveys (CSUR)}, 44(4):1--41, 2012.

\bibitem{bonneau20120science}
Joseph Bonneau.
\newblock The science of guessing: Analyzing an anonymized corpus of 70 million
  passwords.
\newblock In {\em Proceedings of the 2012 IEEE Symposium on Security and
  Privacy}. IEEE, 2012.

\bibitem{bonneau2012quest}
Joseph Bonneau, Cormac Herley, Paul~C Van~Oorschot, and Frank Stajano.
\newblock The quest to replace passwords: A framework for comparative
  evaluation of web authentication schemes.
\newblock In {\em 2012 IEEE Symposium on Security and Privacy}. IEEE, 2012.

\bibitem{bonneau2012birthday}
Joseph Bonneau, S{\"o}ren Preibusch, and Ross Anderson.
\newblock A birthday present every eleven wallets? the security of
  customer-chosen banking pins.
\newblock In {\em International Conference on Financial Cryptography and Data
  Security}. Springer, 2012.

\bibitem{buschek2016snapapp}
Daniel Buschek, Fabian Hartmann, Emanuel Von~Zezschwitz, Alexander De~Luca, and
  Florian Alt.
\newblock Snapapp: Reducing authentication overhead with a time-constrained
  fast unlock option.
\newblock In {\em Proceedings of the 2016 CHI Conference on Human Factors in
  Computing Systems}, 2016.

\bibitem{casimiro2020quest}
Maria Casimiro, Joe Segel, Lewei Li, Yigeng Wang, and Lorrie~Faith Cranor.
\newblock A quest for inspiration: How users create and reuse pins.
\newblock In {\em Adventures in Authentication Workshop}, 2020.

\bibitem{chiasson2006usability}
Sonia Chiasson, Paul~C van Oorschot, and Robert Biddle.
\newblock A usability study and critique of two password managers.
\newblock In {\em USENIX Security Symposium}, 2006.

\bibitem{de2014now}
Alexander De~Luca, Marian Harbach, Emanuel von Zezschwitz, Max-Emanuel Maurer,
  Bernhard~Ewald Slawik, Heinrich Hussmann, and Matthew Smith.
\newblock Now you see me, now you don't: protecting smartphone authentication
  from shoulder surfers.
\newblock In {\em Proceedings of the SIGCHI Conference on Human Factors in
  Computing Systems}, 2014.

\bibitem{de2010colorpin}
Alexander De~Luca, Katja Hertzschuch, and Heinrich Hussmann.
\newblock Colorpin: securing pin entry through indirect input.
\newblock In {\em Proceedings of the SIGCHI Conference on Human Factors in
  Computing Systems}, 2010.

\bibitem{deluca2010atm-sec}
Alexander De~Luca, Marc Langheinrich, and Heinrich Hussmann.
\newblock Towards understanding {ATM} security: A field study of real world
  {ATM} use.
\newblock In {\em Proceedings of the Sixth Symposium on Usable Privacy and
  Security}, SOUPS. ACM, 2010.

\bibitem{de2013back}
Alexander De~Luca, Emanuel Von~Zezschwitz, Ngo Dieu~Huong Nguyen, Max-Emanuel
  Maurer, Elisa Rubegni, Marcello~Paolo Scipioni, and Marc Langheinrich.
\newblock Back-of-device authentication on smartphones.
\newblock In {\em Proceedings of the SIGCHI Conference on Human Factors in
  Computing Systems}, 2013.

\bibitem{dhamija2000deja}
Rachna Dhamija and Adrian Perrig.
\newblock Deja vu-a user study: Using images for authentication.
\newblock In {\em USENIX Security Symposium}, 2000.

\bibitem{egelman2014you}
Serge Egelman, Sakshi Jain, Rebecca~S Portnoff, Kerwell Liao, Sunny Consolvo,
  and David Wagner.
\newblock Are you ready to lock?
\newblock In {\em Proceedings of the ACM SIGSAC Conference on Computer and
  Communications Security}. ACM, 2014.

\bibitem{eiband2017understanding}
Malin Eiband, Mohamed Khamis, Emanuel Von~Zezschwitz, Heinrich Hussmann, and
  Florian Alt.
\newblock Understanding shoulder surfing in the wild: Stories from users and
  observers.
\newblock In {\em Proceedings of the 2017 CHI Conference on Human Factors in
  Computing Systems}. ACM, 2017.

\bibitem{foo2010timing}
Denis Foo~Kune and Yongdae Kim.
\newblock Timing attacks on pin input devices.
\newblock In {\em Proceedings of the 17th ACM Conference on Computer and
  Communications Security}, 2010.

\bibitem{garfinkel2014usable}
Simson Garfinkel and Heather~Richter Lipford.
\newblock Usable security: History, themes, and challenges.
\newblock {\em Synthesis Lectures on Information Security, Privacy, and Trust},
  5(2), 2014.

\bibitem{grand_view_smartlock}
{Grand View Research}.
\newblock {Smart Lock Market Demand To Reach 34.9 Million Units By 2027}.
\newblock
  \url{https://www.grandviewresearch.com/press-release/global-smart-lock-market},
  2020.
\newblock Last accessed June, 2020.

\bibitem{harbach2014sa}
Marian Harbach, Emanuel Von~Zezschwitz, Andreas Fichtner, Alexander De~Luca,
  and Matthew Smith.
\newblock It’sa hard lock life: A field study of smartphone (un) locking
  behavior and risk perception.
\newblock In {\em 10th Symposium On Usable Privacy and Security (SOUPS)}, 2014.

\bibitem{hayashi2011diary}
Eiji Hayashi and Jason Hong.
\newblock A diary study of password usage in daily life.
\newblock In {\em Proceedings of the SIGCHI Conference on Human Factors in
  Computing Systems}. ACM, 2011.

\bibitem{holm1979simple}
Sture Holm.
\newblock A simple sequentially rejective multiple test procedure.
\newblock {\em Scandinavian Journal of Statistics}, pages 65--70, 1979.

\bibitem{huh2015memorability}
Jun~Ho Huh, Hyoungshick Kim, Rakesh~B Bobba, Masooda~N Bashir, and Konstantin
  Beznosov.
\newblock On the memorability of system-generated pins: Can chunking help?
\newblock In {\em Eleventh Symposium On Usable Privacy and Security (SOUPS)},
  2015.

\bibitem{jakobsson2011bootstrapping}
Markus Jakobsson and Debin Liu.
\newblock Bootstrapping mobile pins using passwords.
\newblock
  \url{http://www.markus-jakobsson.com/wp-content/uploads/W2SP11-JL.pdf}, 2011.

\bibitem{kaye2011self}
Joseph'Jofish' Kaye.
\newblock Self-reported password sharing strategies.
\newblock In {\em Proceedings of the SIGCHI Conference on Human Factors in
  Computing Systems}. ACM, 2011.

\bibitem{khan2018evaluating}
Hassan Khan, Urs Hengartner, and Daniel Vogel.
\newblock Evaluating attack and defense strategies for smartphone pin shoulder
  surfing.
\newblock In {\em Proceedings of the 2018 CHI Conference on Human Factors in
  Computing Systems}. ACM, 2018.

\bibitem{kim2012pin}
Hyoungshick Kim and Jun~Ho Huh.
\newblock Pin selection policies: Are they really effective?
\newblock {\em Computers \& Security}, 31(4):484--496, 2012.

\bibitem{kruskal1952use}
William~H. Kruskal and W.~Allen Wallis.
\newblock Use of ranks in one-criterion variance analysis.
\newblock {\em Journal of the American Statistical Association},
  47(260):583--621, 1952.

\bibitem{leiva2014bod}
Luis~A Leiva and Alejandro Catal{\`a}.
\newblock Bod taps: an improved back-of-device authentication technique on
  smartphones.
\newblock In {\em Proceedings of the 16th International Conference on
  Human-Computer Interaction with Mobile Devices \& Services}, 2014.

\bibitem{macorr_loyalty}
{Macorr Research Blog}.
\newblock {Customer Loyalty Cards in Canada}.
\newblock \url{http://www.macorr.com/blog/?p=342}, 2017.
\newblock Last accessed June, 2020.

\bibitem{markert-20-pin-blacklist}
Philipp Markert, Daniel~V. Bailey, Maximilian Golla, Markus D\"{u}rmuth, and
  Adam~J. Aviv.
\newblock {This PIN Can Be Easily Guessed: Analyzing the Security of Smartphone
  Unlock PINs}.
\newblock In {\em IEEE Symposium on Security and Privacy}, SP~'20, pages
  1525--1542, San Francisco, California, USA, May 2020. IEEE.

\bibitem{matthews2016she}
Tara Matthews, Kerwell Liao, Anna Turner, Marianne Berkovich, Robert Reeder,
  and Sunny Consolvo.
\newblock " she'll just grab any device that's closer" a study of everyday
  device \& account sharing in households.
\newblock In {\em Proceedings of the 2016 CHI Conference on Human Factors in
  Computing Systems}, 2016.

\bibitem{win10pin}
{Microsoft}.
\newblock {Passwordless Strategy}.
\newblock
  \url{https://docs.microsoft.com/en-us/windows/security/identity-protection/hello-for-business/passwordless-strategy},
  2018.
\newblock Last accessed June, 2020.

\bibitem{obadaburden2020burden}
Borke Obada-Obieh, Yue Huang, and Konstantin Beznosov.
\newblock The burden of ending online account sharing.
\newblock In {\em Proceedings of the 2020 CHI Conference on Human Factors in
  Computing Systems}. ACM, 2020.

\bibitem{park2018share}
Cheul~Young Park, Cori Faklaris, Siyan Zhao, Alex Sciuto, Laura Dabbish, and
  Jason Hong.
\newblock Share and share alike? an exploration of secure behaviors in romantic
  relationships.
\newblock In {\em Fourteenth Symposium on Usable Privacy and Security (SOUPS)},
  2018.

\bibitem{paypal_pin}
{PayPal}.
\newblock {Password and PIN Security}.
\newblock
  \url{https://www.paypal.com/us/webapps/mpp/security/secure-passwords}, 2018.
\newblock Last accessed June, 2020.

\bibitem{renaud2004my}
Karen Renaud and Antonella De~Angeli.
\newblock My password is here! an investigation into visuo-spatial
  authentication mechanisms.
\newblock {\em Interacting with Computers}, 16(6):1017--1041, 2004.

\bibitem{renaud2015management}
Karen Renaud and Melanie Volkamer.
\newblock Exploring mental models underlying pin management strategies.
\newblock In {\em 2015 World Congress on Internet Security (WorldCIS)}, pages
  18--23, 10 2015.

\bibitem{schechter2015learning}
Stuart Schechter and Joseph Bonneau.
\newblock Learning assigned secrets for unlocking mobile devices.
\newblock In {\em Eleventh Symposium On Usable Privacy and Security (SOUPS)},
  2015.

\bibitem{singh2007password}
Supriya Singh, Anuja Cabraal, Catherine Demosthenous, Gunela Astbrink, and
  Michele Furlong.
\newblock Password sharing: implications for security design based on social
  practice.
\newblock In {\em Proceedings of the SIGCHI Conference on Human Factors in
  Computing Systems}, 2007.

\bibitem{stanekova2013analysis}
L'ubica Stanekov{\'a} and Martin Stanek.
\newblock Analysis of dictionary methods for pin selection.
\newblock {\em Computers \& Security}, 39:289--298, 2013.

\bibitem{stobert2014password}
Elizabeth Stobert and Robert Biddle.
\newblock The password life cycle: user behaviour in managing passwords.
\newblock In {\em 10th Symposium On Usable Privacy and Security (SOUPS)}, 2014.

\bibitem{von2015swipin}
Emanuel Von~Zezschwitz, Alexander De~Luca, Bruno Brunkow, and Heinrich
  Hussmann.
\newblock Swipin: Fast and secure pin-entry on smartphones.
\newblock In {\em Proceedings of the 33rd Annual ACM Conference on Human
  Factors in Computing Systems}, 2015.

\bibitem{von2013survival}
Emanuel Von~Zezschwitz, Alexander De~Luca, and Heinrich Hussmann.
\newblock Survival of the shortest: A retrospective analysis of influencing
  factors on password composition.
\newblock In {\em IFIP Conference on Human-Computer Interaction}. Springer,
  2013.

\bibitem{wang2017understanding}
Ding Wang, Qianchen Gu, Xinyi Huang, and Ping Wang.
\newblock Understanding human-chosen pins: characteristics, distribution and
  security.
\newblock In {\em Proceedings of the 2017 ACM on Asia Conference on Computer
  and Communications Security}. ACM, 2017.

\bibitem{xu2012taplogger}
Zhi Xu, Kun Bai, and Sencun Zhu.
\newblock Taplogger: Inferring user inputs on smartphone touchscreens using
  on-board motion sensors.
\newblock In {\em Proceedings of the 5th ACM Conference on Security and Privacy
  in Wireless and Mobile Networks}, 2012.

\end{thebibliography}
